# A large collection of bioinformatics question-query pairs over federated knowledge graphs: methodology and applications

Jerven Bolleman[1,*], Vincent Emonet[1,*], Adrian Altenhoff[1], Amos Bairoch[1], Marie-Claude Blatter[1], Alan Bridge[1], Severine Duvaud[1], Elisabeth Gasteiger[1], Dmitry Kuznetsov[1], Sébastien Moretti[1], Pierre-Andre Michel[1], Anne Morgat[1], Marco Pagni[1], Nicole Redaschi[1], Monique Zahn-Zabal[1], Tarcisio Mendes de Farias[1,†] and Ana Claudia Sima[1,†]

[1] *SIB Swiss Institute of Bioinformatics, Switzerland*
*Correspondence address: Tarcisio Mendes de Farias, Ana Claudia Sima, SIB Swiss Institute of Bioinformatics, Amphipôle, Quartier UNIL-Sorge, 1015 Lausanne, Switzerland e-mail:*
[tarcisio.mendes@sib.swiss](mailto:tarcisio.mendes@sib.swiss), [ana-claudia.sima@sib.swiss](mailto:ana-claudia.sima@sib.swiss)

## Abstract

**Background**. In the last decades, several life science resources have structured data using the same framework and made these accessible using the same query language to facilitate interoperability. Knowledge graphs have seen increased adoption in bioinformatics due to their advantages for representing data in a generic graph format. For example, yummydata.org catalogs more than 60 knowledge graphs accessible through SPARQL, a technical query language. Although SPARQL allows powerful, expressive queries, even across physically distributed knowledge graphs, formulating such queries is a challenge for most users. Therefore, to guide users in retrieving the relevant data, many of these resources provide representative examples. These examples can also be an important source of information for machine learning, if a sufficiently large number of examples are provided and published in a common, machine-readable and standardized format across different resources.

**Findings**. We introduce a large collection of human-written natural language questions and their corresponding SPARQL queries over federated bioinformatics knowledge graphs (KGs) collected for several years across different research groups at the SIB Swiss Institute of Bioinformatics. The collection comprises more than 1000 example questions and queries, including 65 federated queries. We propose a methodology to uniformly represent the examples with minimal metadata, based on existing standards. Furthermore, we introduce an extensive set of open-source applications, including query graph visualizations and smart query editors, easily reusable by KG maintainers who adopt the proposed methodology.

**Conclusions**. We encourage the community to adopt and extend the proposed methodology, towards richer KG metadata and improved Semantic Web services.

**URL:** https://github.com/sib-swiss/sparql-examples

*co-first authors

†co-last authors

# Introduction

The accuracy and real-world applicability of question answering systems over structured data crucially depend on the availability of high-quality, representative question-query pairs over diverse datasets. Resources such as UniProt (UniProt Consortium, 2023), have been used in bioinformatics for many years, where they facilitate data integration and exploration through (potentially federated) SPARQL queries. However, a long-standing challenge is making this wealth of data usable by a wider range of users, including researchers without technical training. Writing SPARQL queries is outside the competence of most life scientists, as it requires both knowledge of the query language itself and understanding of the knowledge graph schema, for example, a Resource Description Framework (RDF) graph data model. Many solutions have thus been investigated, including Natural Language Interfaces (NLI) to knowledge graphs (KG). The recent progress in Large Language Models (LLMs) has enabled to boost performance across a wide range of Natural Language Processing tasks, including promising results in LLM-based Knowledge Graph Question Answering (KGQA) systems (A.-C. Sima & de Farias, 2023). A recent study highlights that unified models, trained across diverse datasets stemming from distinct knowledge graphs, eliminate the need for separate models for each KG and can therefore significantly reduce the overall costs of fine-tuning and deploying such models (Zahera et al., n.d.). Therefore, a key requirement to the development of novel, LLM-based KGQA systems, is the availability of high-quality, representative public datasets of questions and corresponding SPARQL queries across a wealth of diverse KGs.

Capturing user intent and translating questions becomes significantly more complex when considering multiple, distributed scientific KGs such as those comprising the SIB Swiss Institute of Bioinformatics Semantic Web of Data. Although federated SPARQL queries are very powerful because they enable joint querying of physically distributed knowledge graphs, the challenge lies in formulating such queries. Federated queries are inherently complex because they require users to not only understand the data models of the individual sources involved in the query, but also to be aware of how these sources interconnect (*i.e.*, how they can be joined). Hence, the availability of examples to guide users - or of a system supporting the automatic translation of user questions to federated queries - are critical. However, federated queries have been mostly neglected so far in existing collections of natural language questions and equivalent SPARQL queries.

The SIB hosts a number of high-quality, curated and interoperable life science knowledge graphs[1] that are accessible via public SPARQL endpoints, i.e. web addresses that can receive and process SPARQL queries. Each of these KGs is independently managed by different research groups within the institute. However, a common point across all resources is that they each maintain a set of representative questions and their equivalent SPARQL queries, collected over time. These are made available to inspire and help users formulate new queries over the respective KG data. Importantly, the resources also include examples of federated queries, which allow for joining information from their KG with those of other, complementary KGs, from

---

the SIB catalog and beyond (e.g., Wikidata (Vrandečić & Krötzsch, 2014), IDSM (Galgonek & Vondrášek, 2021), ChEMBL (Zdrazil et al., 2024)). Prior to the work presented in this paper, the examples were provided in heterogeneous and sometimes non-machine-readable formats (e.g., as plain text in a static webpage). However, to facilitate interoperability and more generally the reusability of all examples as a large, uniform collection, it is of paramount importance to standardize how queries are represented, in agreement with the FAIR data principles (Findable, Accessible, Interoperable and Reusable), allowing their automatic discoverability by both users and services.

In this paper, we introduce the collection of over 1,000 human-written, real-world question-query pairs, including 65 federated queries (i.e., queries that ask questions requiring at least 2 sources jointly), collected over several years across 11 SIB interoperable knowledge graphs. Additionally, we describe the methodology we have applied across SIB to represent, store and test example queries by adopting different (meta)data standards. Importantly, we show that this standardization enables easily finding examples across different endpoints, as well as setting up, running, and maintaining services that use these examples. These services include powerful user-facing applications, such as search engines or automatically generated web visualizations. The rapid development and uptake of these services across different SIB Resources directly prove the value of adopting FAIR data principles and metadata standardization. To our knowledge, ours is the most comprehensive portfolio of bioinformatics examples and services facilitated by FAIR KG (meta)data across distributed KGs. Both the collection of example questions-SPARQL queries, as well as all the developed services leveraging these examples, are available open-source and can be adapted to other KGs by following our recommended metadata format.

We encourage the community to engage with the collection of examples, adopt and further extend our proposed format for describing SPARQL examples, and finally reuse and improve the services presented in this work. Efforts towards standardization will not only contribute towards FAIRer Knowledge Graphs, but will also help build a solid foundation for a new generation of Semantic Web Technologies, such as LLM-based Natural Language Interfaces over Knowledge Graphs.

## Sources

At the time of writing, we have collected contributions from 10 large-scale KGs across different SIB groups. Additionally, we integrated contributions from one external KG, "dbgi", with examples contributed by external collaborators in the Digital Botanical Gardens Initiative[2].

For details on the SPARQL endpoints maintained by the SIB, we refer the reader to the SIB Semantic Web of Data (SIB Swiss Institute of Bioinformatics RDF Group Members, 2024). Among the contributing KGs, we can mention UniProt - currently the largest publicly available

---

[2] https://www.dbgi.org/

KG - with over 168B triples and a total of 61 contributed example queries. To our knowledge, the SIB collection is the most comprehensive collection of real-world SPARQL queries examples to-date in bioinformatics.

As an estimate of the complexity of the queries in this collection, we show the average number of triple patterns (TP) per endpoint included in the collection in Figure 1. The number of TPs is a good approximation for the complexity of a query, because it indicates the number of hops (joins) required to compute an answer over the graph. Most queries have around 6 TPs on average, with the notable exception of the DBGI, for which queries have a much higher complexity, owing to the complexity of the data model, but also due to federation with Wikidata, which is present in most examples. On the other end, HAMAP includes only a few very simple examples, as it is a small KG (only 7 classes) storing annotation rules. We note that most existing large scale collections focus on simple queries, i.e. queries with at most 3 TPs (for examples, see Related Work).

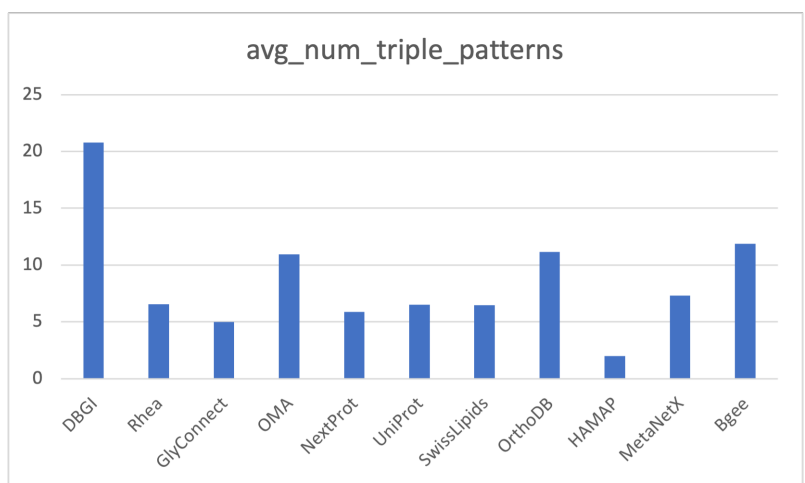

Figure 1. Average number of triple patterns per query in each of the KGs contributing to the examples collection

## Methodology

Standardization is at the core of our approach. To structure and describe the question-query pairs, we rely on the SHapes Constraint Language (SHACL) which is RDF based[3]. In addition to SHACL and RDF Schema (RDFS)[4], we describe and annotate the examples with Schema.org. Standardizing how we represent the examples is crucial since they can target different and independent data sources managed by distinct groups and organizations. A standardized, machine-readable representation, enables both users and services to discover and smoothly use the question-query descriptions that are defined in RDF and stored in distributed data stores accessible with SPARQL. More precisely, example question-query pairs can be stored with their associated metadata in the same RDF store, consequently, making them accessible via the same SPARQL endpoint. As a result, this standardization and infrastructure facilitate the automatic testing of queries, as well as the deployment of services with minimal code modifications across endpoints, such as the generation of user-friendly Web pages with query descriptions and graphical visualizations (see Section Applications).

---

[3] https://www.w3.org/TR/shacl/
[4] https://www.w3.org/TR/rdf-schema/

Figure 2 illustrates the end-to-end workflow, starting from examples contributed by KG maintainers to services consuming the examples from the named graph they are published in.

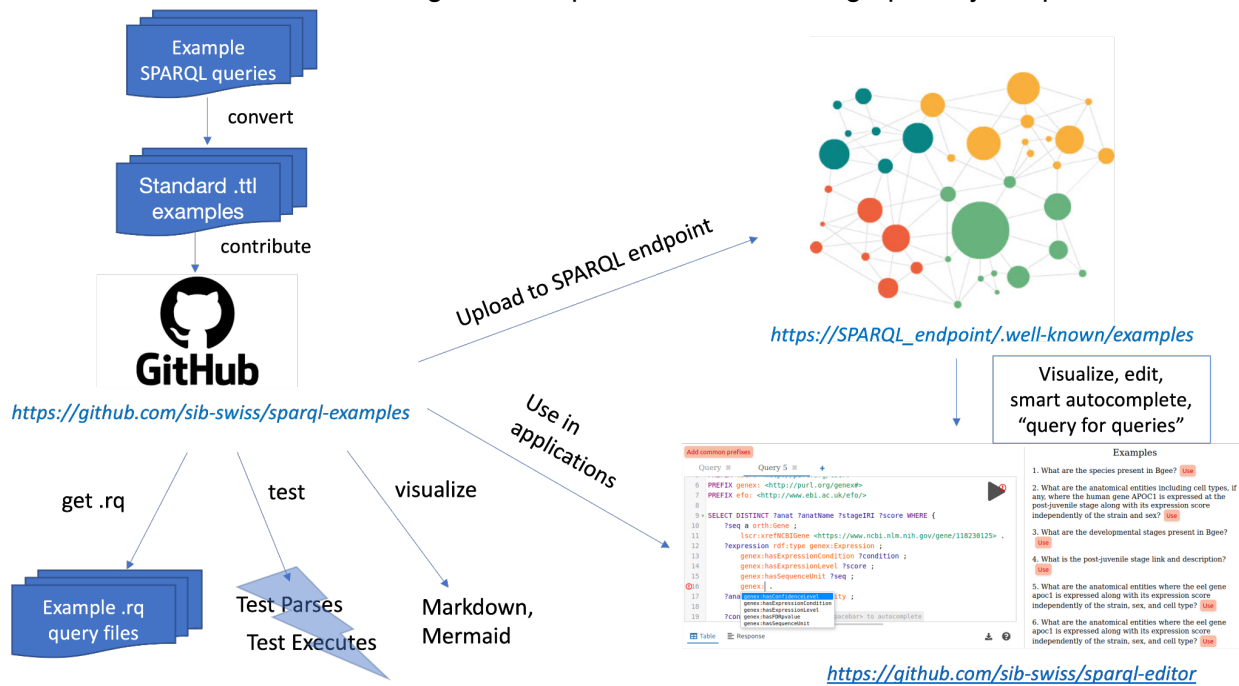

Figure 2. End-to-end workflow, from example contributions to services that uniformly consume examples across distinct SPARQL endpoint (e.g. sparql-editor and Bio-Query template search)

## Implementation

We collaborate in an open GitHub repository (https://github.com/sib-swiss/sparql-examples). The repository is the "single source of truth", where all SPARQL examples are collected via contributions from data store maintainers. Each example is stored in a separate file using the Terse RDF Triple Language (Turtle, *i.e.* ".`ttl`") syntax and file format[5], within a folder named according to the resource that contributed the example. For example, all Bgee queries can be found in https://github.com/sib-swiss/sparql-examples/tree/master/examples/Bgee .

Each Turtle file stores a single example, which is represented by:
  a) The type of the query (e.g., Ask, Select, Describe).
  b) The description of the query (e.g., the question in plain text), including a language tag.
  c) The SPARQL query itself.
  d) The target SPARQL endpoint(s) where the query can be executed.

---

[5] https://www.w3.org/TR/turtle/

e) Additional keywords (tags).

Each question-query pair is defined as a *SPARQLExecutable* SHACL[6] concept. We also specify the *SPARQLExecutable* subtype, such as *SPARQLAskExecutable* and *SPARQLSelectExecutable*, that is executable queries based on an *ASK* and *SELECT SPARQL* query, respectively. Furthermore, all of the fields b) to e) above are represented in a structured, standardized and machine-readable format. Namely, the plain English question or description of the query can be described via an *rdfs:comment,* the query itself via *sh:select, sh:ask, sh:update, sh:construct* or *sparql-examples:describe*[7] (according to the type of the SPARQL query, *i.e.* ASK, SELECT etc.). For multilingual support in the examples, all questions are annotated with a corresponding language tag (e.g. "`@en`" for English). The set of queries in this work only includes Ask, Select and Describe queries. As DESCRIBE queries are defined in SPARQL 1.1 and can not be used in SHACL, we introduce a new property and concept to represent this type of query, that is, *sparql-examples:describe* where the *sparql-examples:* prefix corresponds to *https://purl.expasy.org/sparql-examples/ontology#*. The target SPARQL endpoint(s) are described with *schema:target*. Note that there can be several such endpoints where the query is directly executable. The query reusability (without any modifications) stems from the adoption of the FAIR principles in the design of KGs across the entire SIB Semantic Web of Data, which makes them uniformly queryable when requesting the same type of information. For example, the query defined at https://www.bgee.org/sparql/.well-known/sparql-examples/1 is usable as-is in at least three SPARQL endpoints, namely Bgee, OMA and UniProt. In the case of this example, it retrieves the list of species in each KG. The *schema:target* shows where users can run these queries and also demonstrates the benefit of sharing knowledge representation of common concepts and values. Finally, example queries can optionally be annotated with additional keywords (tags) using *schema:keywords* property.

To further illustrate how question-query pairs are structured, Listing 1 shows another example of a SPARQL query that retrieves all taxa in UniProt and its corresponding question in plain English. The SPARQL query is defined using the *sh:select* relation, while the corresponding natural language question (i.e., the question in plain English that corresponds to the query) is always assigned to the *rdfs:comment* relation.

```
PREFIX ex: <https://sparql.uniprot.org/.well-known/sparql-examples/>
PREFIX sh: <http://www.w3.org/ns/shacl#>
PREFIX rdf: <http://www.w3.org/1999/02/22-rdf-syntax-ns#>
PREFIX rdfs:<http://www.w3.org/2000/01/rdf-schema#>

ex:001 a sh:SPARQLSelectExecutable, sh:SPARQLExecutable ;
    sh:prefixes _:sparql_examples_prefixes ;
    rdfs:comment "Select all taxa from the UniProt taxonomy"@en ;
```

---

[6] https://w3c.github.io/shacl/shacl-sparql/#dfn-sparql-based-constraint
[7] https://purl.expasy.org/sparql-examples/ontology#SPARQLDescribeExecutable

```
    sh:select """PREFIX up: <http://purl.uniprot.org/core/>
SELECT ?taxon
FROM <http://sparql.uniprot.org/taxonomy>
WHERE
{
    ?taxon a up:Taxon .
}""" ;
    schema:target <https://sparql.uniprot.org/sparql/> ;
    schema:keywords "taxa" .
```

Listing 1. Question-query pair to select all UniProt taxa. We define this as a *sh:SPARQLSelectExecutable* where the question is assigned to *rdfs:comment* (with an English language tag annotation), while its corresponding SPARQL query in *sh:select*. The *schema:target* relation is used to define the target SPARQL endpoint where the query is executable. Optionally, the *sh:prefixes* relation is also stated to report all prefixes needed to run the query. This example has been uploaded to the UniProt SPARQL endpoint and can be viewed and executed at https://sparql.uniprot.org/.well-known/sparql-examples/1.

After describing the SPARQL examples, we also store them in a standardized location directly as a subgraph within the KG that they target - that is, as part of the same SPARQL endpoint as the data they act on - so as to enhance their findability. This allows both users and services to uniformly retrieve examples of interest, or to aggregate information about the collection of examples available in each resource, by directly querying the SPARQL endpoints themselves. In addition, it enables reusability of services across KGs with minimal modifications and without reliance on a centralized repository. We provide a script to merge the examples targeting a given endpoint as a single *.ttl* file that can be uploaded to a dedicated subgraph (e.g., as a named graph) within each KG. Across SIB, we have chosen to upload examples to a named graph composed of the SPARQL endpoint web address and the *".well-known/sparql-examples"* suffix. Examples can then be assigned a unique identifier, more precisely, an Internationalized Resource Identifier (IRI), which may also be resolvable, *i.e.* allow access to the query using standard HTTP(S) for resource maintainers that want to associate specific webpages to the queries. For example, the query that retrieves all taxa over the UniProt SPARQL endpoint can be accessed at https://sparql.uniprot.org/.well-known/sparql-examples/1.

Once uploaded in the SPARQL endpoint, the examples themselves can be processed (for example, to search for relevant examples on a given topic) through simple SPARQL queries. Listing 2 shows how to retrieve example question-query pairs where the question contains the keyword "*species*" from a given endpoint. The following query can be directly executed, for example, over the UniProt[8] or the Bgee[9] SPARQL endpoint:

```
PREFIX rdfs: <http://www.w3.org/2000/01/rdf-schema#>
```

---

```
PREFIX sh: <http://www.w3.org/ns/shacl#>
SELECT * WHERE {
  ?ex sh:select ?query .
  ?ex rdfs:comment ?question.
  FILTER (contains(?question, "species")) }
```

Listing 2. Query to retrieve all examples that include questions containing the keyword "*species*" from a SPARQL endpoint where formatted SPARQL examples have been uploaded

Finally, the landing page of our open GitHub repo provides an extensive description of how to apply our methodology to other KGs (RDF data stores). We include instructions and automated scripts to help describe new examples (i.e., query-question pairs), as well as explaining where such examples should be stored, how they can be queried (*i.e.,* "querying for queries"), tested automatically etc. To help KG maintainers check whether their endpoint meets all criteria required for metadata annotation, we have also developed an automated checker available online[10]. The checker solely requires a SPARQL endpoint URL to validate the existence of properly formatted metadata, such as SPARQL examples and the Vocabulary of Interlinked Datasets (VoID)[11] description, in the given endpoint. For endpoints which do not have the full metadata available, the checker suggests concrete steps that can be taken, for example, to add the corresponding VoID description to the endpoint by using our open-source VoID generator[12].

# Applications

In this section, we provide several applications facilitated by the collection of example queries, such as:

- Automated query testing across SIB endpoints.
- Uniform query search and editing with a reusable SPARQL editor.
- Automatically generating graph visualizations of queries.
- Searching for scientific questions in the Bio-Query template interface.

All the applications can be deployed across the entire collection of questions and queries, owing to the standardized representation of examples that we have designed. The applications can be reused over any question-query collections that follow the approach described in the "Methodology" Section.

## Tooling to help maintain SPARQL query examples

To help teams maintain their own SPARQL query examples, we provide a set of tools in a separate GitHub repository[13]. This includes testing for continuous integration, visualizations to help understand the queries, and a set of automated fixes that can remove common errors. We provide details on each of these components in the following sections.

---

[10] https://sib-swiss.github.io/sparql-editor/check
[11] https://www.w3.org/TR/void/
[12] https://github.com/JervenBolleman/void-generator
[13] https://github.com/sib-swiss/sparql-examples-utils

## Automated testing

Each SPARQL query is tested for syntactical correctness using two different parsers: the first provided by the Apache Jena[14] project, and a second by the eclipse RDF4j[15] project. The metadata about the query is tested using a set of SHACL rules with the RDF4j implementation of SHACL. These rules check if the structure of the metadata around the query matches the expected format, avoiding errors in the contributed examples. For example, the presence of all mandatory fields (query ID, query type, comment, target) is checked. Furthermore, the rules check if the query matches the metadata—for example, that all queries declared to be federated queries actually contain a SERVICE clause in the query text.

Optionally, we can also execute all queries on their remote target endpoints. We chose to only check that the queries retrieve at least one result, by programmatically changing the queries. More specifically, we add a limit which restricts the queries to only retrieve one result (i.e., a LIMIT[16] clause with a value of 1). This avoids introducing non-essential load on the tested SPARQL servers, since some queries would result in long-running times or high computational load. Finally, all SPARQL endpoints used in federated queries are tested to be alive and responsive to SPARQL queries. These tests are written in Java using JUnit 5 Jupiter[17] as the test framework. Tests that do not require remote SPARQL endpoints are deployed as GitHub actions that run on each push to the shared git repository, ensuring the validity of newly contributed queries. Tests that require network or that access the remote SPARQL endpoints are only run on demand by the query writers and can be used as regression tests when changes are made to the target repositories (i.e., to ensure that existing queries are still valid). Many of the query examples are relatively complex and would therefore add significant traffic to the remote endpoints if executed frequently. We thus only run these tests on demand.

## Enabling query visualization

Complex SPARQL queries are difficult to follow for the majority of users. To assist in understanding the existing question-query examples, we automatically generate Markdown[18] files with visual graphs and descriptions for each example query. To visualize these, we use

---

Mermaid[19], a diagramming and charting tool inspired on the Markdown text syntax. Namely, we automatically generate for each example query a corresponding visual graph in Mermaid. The generated markdown files can be displayed as GitHub pages[20] or as Web pages in general (i.e., rendered HTML[21]). These Web pages can be then accessible, for example, via the web address used as their IRI. To facilitate the findability and understanding of the examples, we make the full collection, including Mermaid visualizations for each query, available online in GitHub Pages[22]. Figure 3 shows a simple Mermaid-based graph corresponding to an example query that asks for disease related proteins located inside the cell. This query can be executed at the UniProt SPARQL endpoint.

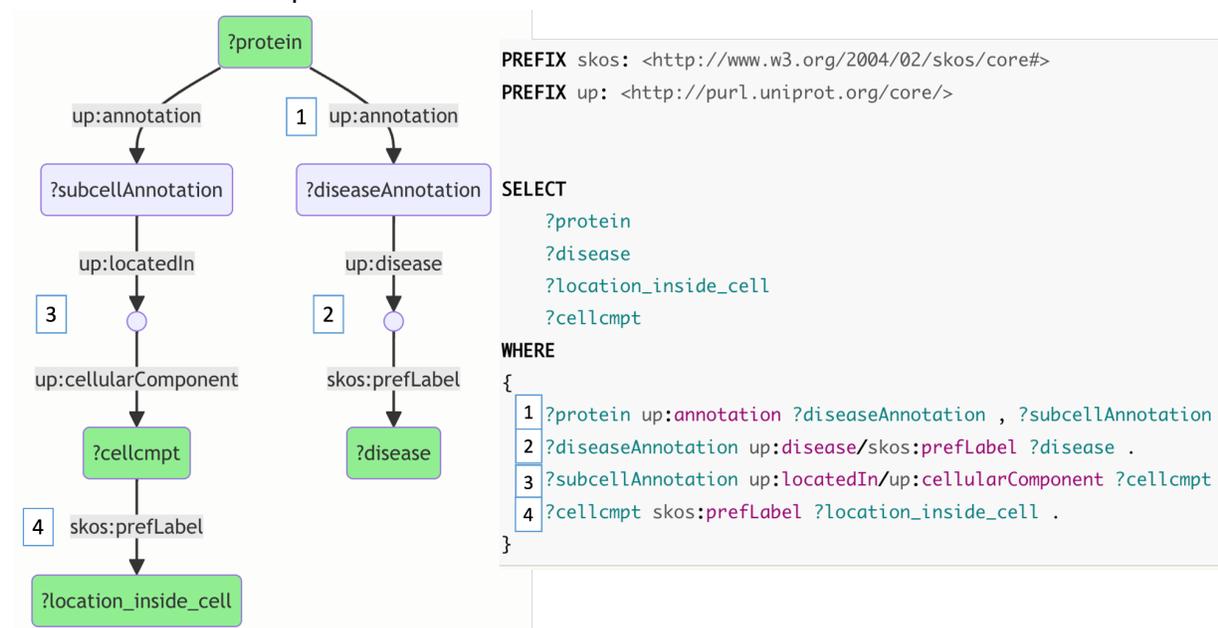

Figure 3. A graphical representation of [UniProt example query 21](#) retrieving disease related proteins that are known to be located within the cell. This visualization is available in the GitHub pages of the collection of SPARQL examples [here](#). In the figure, green cells show the projected variables, i.e. the variables that are selected to be included in the final result. The small circles act as intermediary, undefined variables (i.e., intermediary blank nodes) introduced by decomposing graph paths. The gray boxes shown as edge labels are the properties in the query triple patterns. In this figure, the numbered boxes shown on the left side of edges help to match the visualization to the corresponding triple patterns in the query shown on the right side.

## Fixing query examples to be fully compliant with SPARQL 1.1

There are a number of query engines that have extended SPARQL 1.1 in non-standard ways. We implemented a module that attempts to "fix" these queries to be fully compliant with the SPARQL 1.1 specification. For example, Anzo and Blazegraph have implemented an extension for named subqueries. This named subquery extension is widely used in the Wikidata community. Nevertheless, these queries can often be semantically rewritten by following the

---

SPARQL 1.1 specification. More precisely, our "fixer" approach parses the queries with the Blazegraph SPARQL parser and browses specific abstract syntax trees to look for such named subqueries and replace them with SPARQL 1.1 standard-compliant subqueries.

Another common mistake is that query examples might miss the prefix declaration required to make a query run. We store the common prefixes used in our resources, with an option for project specific prefixes, we can easily add missing ones to a query without further user interaction.

## Query editing and searching

We extend the widely used Yasgui SPARQL environment (Rietveld & Hoekstra, 2013) with a new section showing the queries in use. The code for this extension is available as a package at the *npm* software registry[23] and can be easily deployed by SPARQL endpoint maintainers using a single custom HTML element `<sparql-editor/>`. We welcome contributions to develop this further at https://github.com/sib-swiss/sparql-editor. The editor[24] includes several useful features. It automatically pulls query examples from the endpoint and presents them to the user alongside the editor, so users can easily reuse these queries to write their own by editing or extending them. It also provides a precise autocomplete based on the VoID descriptions that summarize an endpoint content. Whenever available, these descriptions are retrieved by the editor that queries the endpoint itself. The main information of interest processed by the editor from the VoID descriptions are the classes and properties that connect them. The VoID description is directly extracted from the SPARQL endpoint that is used by our extended Yasgui environment, therefore accurately reflecting the actual data. This ensures that only properties that are indeed stated, applicable according to the schema and the known type of a variable are suggested in the autocomplete dropdown menu. For further details on the VoID description generator used, see the corresponding GitHub repository[25].

## Integration in Bio-Query

The standardized representation of example queries facilitates their use in applications downstream. As an example, we have also incorporated the collection as-is in the Bio-Query (A. C. Sima et al., 2019) interface available online[26]. Figure 4 shows an example question and query from the collection. The questions are automatically assigned to categories named according to the resource that contributed them. The collection can be found under the "*SIB Example SPARQL queries*" category in the public Bio-Query interface.

---

[23] https://www.npmjs.com/package/@sib-swiss/sparql-editor
[24] https://sib-swiss.github.io/sparql-editor
[25] https://github.com/JervenBolleman/void-generator
[26] https://biosoda.expasy.org/build_biosodafrontend/

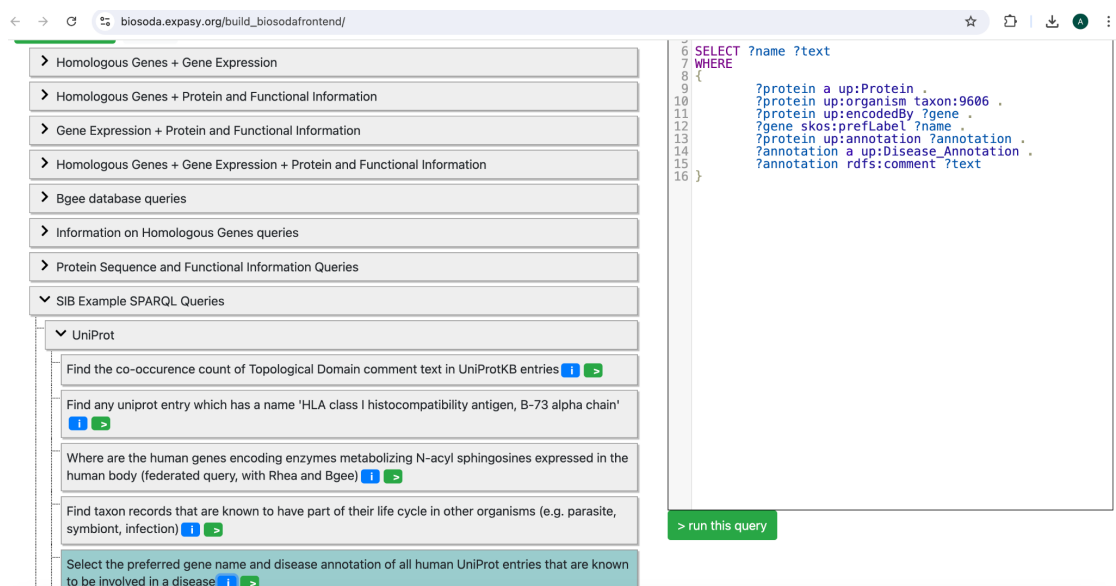

Figure 4. Example UniProt entry in the Bio-Query interface, integrated from the collection of SIB Example SPARQL queries. The collection is processed from the central GitHub repo in order to produce the JSON representation required by the interface. The standardization of the examples minimizes the effort to integrate these in the common interface, which can then act as a central hub to search for, and adapt, existing examples across SIB KGs.

## Challenges and Lessons Learned

Efforts towards standardization are always faced with challenges at multiple levels. We have found that, even with community agreement on a common metadata format, simple details, such as non-uniform HTTP response headers expected across the different endpoints, can still hamper the uniform processing of the metadata. Moreover, even the queries themselves—although in principle portable across different triplestores—are not always reusable across different implementations. This is because, as mentioned previously, some query engines have extended SPARQL 1.1 in non-standard ways. An example is the way that "magic triples" are used to provide query execution hints[27] in Blazegraph or AWS Neptune. These "magic triples" need to be removed when moving to a different SPARQL endpoint, as they would otherwise refer to triples that do not exist.

At a higher-level, one of the biggest challenges we still face is that the collection of questions and queries were designed as generic guidelines for users, and not to be consumed by machines directly. This poses a set of unique challenges also in reusing them for the purpose of training and evaluating machine learning algorithms to automatically translate questions into equivalent SPARQL queries. End-user questions are often more abstract than their corresponding SPARQL queries, because the information searched by the user is not explicitly defined in the question. For example, let us consider the question: "What are the species available?", in this question we do not know what is the exact information the user is looking for.

---

Is the user searching for common names or scientific names of species, or both? Is the user looking for a specific list of species identifiers of a given taxonomy? On the other hand, in a SPARQL query we often define the exact information to be retrieved (i.e., query projection). This impacts the retrieved results and potentially their amount. Therefore, while for an end-user the exact projected variables do not seem to be an essential point in the translation of a question to a query, for automatically evaluating an ML system, the set of projected variables must be precisely defined in the question (e.g., "*return only the species scientific names*", "*return only proteins and their associated gene names*") as opposed to being loosely defined, which would hamper a precise evaluation. Some questions in the catalog have also been designed in sequence, such that the text of one question is a follow-up to a previous one (e.g., "*Same as previous but with P0ABU7 as a query*"). Unless consistently done so throughout the collection, such follow-up questions would need to be discarded or rewritten, given that training a system to understand follow-ups would require a completely different approach.

Finally, an added challenge (common to all collections of question-query pairs) is that the SPARQL endpoints the queries target might change or not even be available anymore, rendering the queries harder to execute or reproduce. We acknowledge that ensuring long-term preservation of the corresponding data is a difficult task, however precisely due to this reason we cannot maintain dedicated copies of the datasets targeted by the collection of queries. Instead, we rely on the resource providers to provide a long-term solution, which can be, for example, automatically redirect the existing SPARQL endpoint link to a site where the data is available at least in an archived format, from where users can deploy it locally.

Nevertheless, in spite of these challenges, the portfolio of reusable services that we have developed and deployed across independently managed SIB endpoints attests to the fact that even small steps towards standardization can have an important impact.

## Related Work

Many existing works have proposed collections of Natural Language questions and corresponding SPARQL queries, usually designed as benchmarks to facilitate the task of automatically translating questions to queries. However, existing collections suffer from shortcomings, which we mention below:

1) Non-federated: to the best of our knowledge, existing peer-reviewed published benchmarks, either large or small scale, do not consider federated queries and are usually centered around a single endpoint. We can mention here the examples of (Dubey et al., 2019; Trivedi et al., 2017). Even multi-dataset collections usually effectively address one single endpoint at a time, such as (Dubey et al., 2019; Kosten et al., 2023; Trivedi et al., 2017). Consequently, existing KGQA systems also focus on simple, non-federated queries (Zafar et al., 2018), often solely over DBPedia or Wikidata (Diefenbach et al., 2018). The existing collections of federated SPARQL queries, such as (Aimonier-Davat et al., 2024; Saleem et al., 2018) are designed to merely evaluate SPARQL federation engines and do not include corresponding natural language questions.

2) Non-representative: crafting a large-scale benchmark manually is an extremely costly endeavor, therefore large collections (see (Dubey et al., 2019)) have been generated semi-automatically, which does not ensure that the questions accurately represent the information needs of real users. In contrast, our collection has been collected over time by different KG maintainers across the SIB, where most examples reflect questions asked by real users of the data.

3) Small scale: the higher-complexity, human-written question-query collections, such as the QALD series (Question Answering over Linked Data challenge updated each year, e.g. (Usbeck et al., 2023)), only include a small number of queries. In contrast, our dataset includes over 1,000 question-query pairs, all of which have been human-written and collected over time from the different KG maintainers.

4) HAMAP as SPARQL (Bolleman et al., 2020) is also a collection of more than 4,000 queries for a specific goal, annotating proteins. However, these are not aimed at educating users to the capabilities of a system. Nor are they annotated with a text for what their purpose is.

5) SPIN[28], SPARQL Inference Notation, is a W3C member submission that informed some of the design decisions of SHACL. SPIN provides for SPARQL templates and has logic for describing arguments to queries. However, its focus on constraints lead its modeling logic to ignore the concept of using these queries to mark up examples.

6) The Wikidata project has a media wiki template[29] for SPARQL queries. However, this does not standardize the metadata about the queries, nor does it provide the necessary utilities to help users maintain their queries.

## Conclusion

In this paper, we introduce a large collection of example questions and their corresponding SPARQL queries collected over time across the catalog of SIB KGs and beyond. Moreover, we have proposed a methodology to standardize the representation of these examples, facilitating their uniform processing by both users and services. Finally, we have presented a comprehensive portfolio of services that leverage the standardization of metadata across the collection of examples, including services for the automated testing, editing and visualizing of example queries. All of these are available open-source and can easily be adapted by other KG maintainers for their own purposes, provided that maintainers adopt our recommended approach.

In the future, we plan to extend the current methodology to support defining templates in the example queries, in other words, fields that can be dynamically filled by querying a dedicated API (e.g., a SPARQL query). This would be a generalization of our current format, avoiding hard-coding literals in the example queries. We also plan to deploy an LLM-based solution to automatically translate user questions in natural language to machine-readable SPARQL queries, leveraging the collection presented here.

---

[28] https://spinrdf.org/
[29] https://www.wikidata.org/wiki/Template:SPARQL

We encourage the community to contribute to our proposed metadata specification, towards FAIRer Knowledge Graphs, which will pave the way for improved Semantic Web Tools. The advent of new technologies, such as Large Language Models, provide an opportunity for enabling a wider audience, including researchers and the public at large, to fully benefit from the wealth of interconnected data available in distributed KGs. We believe that a set of representative metadata standards for KGs, including the SPARQL examples metadata proposed here, represent an important and pragmatic step forward in this direction.

## Competing interests

The authors declare that they have no competing interests.

## Acknowledgements

We acknowledge contributions by all members of the SIB Semantic Web Focus Group, as well as funding from the State Secretariat for Education, Research and Innovation SERI. This work was also supported by the CHIST-ERA grant CHIST-ERA-22-ORD-09 by SNF grant 20CH21_217482; and the Swiss Open Research Data Grants (CHORD) in Open Science I, a program coordinated by swissuniversities, grant id "Swiss DBGI-KM".

UniProt is supported by the Swiss Federal Government through the State Secretariat for Education, Research and Innovation SERI, and by the National Human Genome Research Institute (NHGRI), Office of Director [OD/DPCPSI/ODSS], National Institute of Allergy and Infectious Diseases (NIAID), National Institute on Aging (NIA), National Institute of General Medical Sciences (NIGMS), National Institute of Diabetes and Digestive and Kidney Diseases (NIDDK), National Eye Institute (NEI), National Cancer Institute (NCI), National Heart, Lung, and Blood Institute (NHLBI) of the National Institutes of Health under grant U24HG007822.

## Authors' Contributions

Conceptualization: A.S, J.B., V.E., T.M.F. Software: A.S, J.B., V.E., T.M.F. Investigation: A.S, J.B., V.E., T.M.F. Supervision: A.S, T.M.F. Writing, original draft: A.S, J.B., V.E., T.M.F. Writing, review and editing: all authors. Funding acquisition: A.S, J.B., T.M.F, S.D., Al. B.

## Data Availability

The catalog of examples is available online at https://github.com/sib-swiss/sparql-examples . All the datasets referenced are publicly available, while the services leveraging the examples are all available open-source. The links to each of these are provided in the article text.

## Abbreviations

DBGI - Digital Botanical Gardens Initiative
FAIR - Findable, Accessible, Interoperable, Reusable
HTTP - Hypertext Transfer Protocol
KG - Knowledge Graph
KGQA - Knowledge Graph Question Answering system
LLM - Large Language Model
QALD - Question Answering over Linked Data
RDF - Resource Description Framework
RDFS - RDF Schema
SHACL - Shapes Constraint Language
SIB - Swiss Institute of Bioinformatics
SPARQL - SPARQL Protocol and RDF Query Language
TTL - Terse RDF Triple Language
VoID - Vocabulary of Interlinked Datasets